\documentclass[conference,english,a4paper]{IEEEtran}
\usepackage[T1]{fontenc}
\usepackage[latin1]{inputenc}
\usepackage{graphicx}
\usepackage{amssymb}
\usepackage{amsmath}

\makeatletter

\providecommand{\boldsymbol}[1]{\mbox{\boldmath $#1$}}

\usepackage{babel}
\makeatother

\begin{document}

\title{Distributed Space Time Codes for the Amplify-and-Forward Multiple-Access Relay Channel}

\author{\authorblockN{Maya Badr}\authorblockA{E.N.S.T.\\46, rue Barrault\\75013 Paris, France\\Email: \texttt{mbadr@enst.fr}}\and\authorblockN{Jean-Claude Belfiore}\authorblockA{E.N.S.T.\\46, rue Barrault\\75013 Paris, France\\Email: \texttt{belfiore@enst.fr}}}

\maketitle
\begin{abstract}
In this work, we present a construction of a family of space-time block codes for a Multi-Access Amplify-and-Forward Relay channel with two users and a single half-duplex relay. It is assumed that there is no Channel Side Information at the transmitters and that they are not allowed to cooperate together. Using the Diversity Multiplexing Tradeoff as a tool to evaluate the performance, we prove that the proposed scheme is optimal in some sense. Moreover, we provide numerical results which show that the new scheme outperforms the orthogonal transmission scheme, \textit{e. g.} time sharing and offers a significant gain.
\end{abstract}

\section{Introduction}
In a Multiple Access Relay (MAR) channel, first introduced in \cite{Kramer}, multiple users communicate with a single destination with the help of some relays. Many cooperation protocols were applied to the MAR channel, such as the well-known Dynamic Decode-Forward (DDF) protocol \cite{On-the-opt} and the Compress-Forward (CF) protocol \cite{erkip}. As a performance analysis tool, the Diversity Multiplexing Tradeoff (DMT), first introduced by Zheng and Tse in \cite{Zheng-1}, is used. Namely, the upper bound on the achievable DMT for the MAR channel derived in \cite{On-the-opt} is used to compare different strategies.

In the DDF strategy, both users transmit their information symbols throughout an entire block, the relay decodes the information it receives only when it has a sufficient information for a correct detection. Then the relay re-encodes the message and transmits it to the destination. It is shown in \cite{On-the-opt} that the DDF protocol achieves the optimal DMT for low multiplexing gains while being suboptimal for high multiplexing gains. In the CF protocol, the relay uses source coding to compress its received signal and forward it to the destination. It is shown in \cite{erkip} that this communication strategy achieves the optimal DMT for high multiplexing gains, but suffers from a diversity loss for a low multiplexing gain.

In this work, we consider the Multi-Access Amplify-Forward (MAF) recently proposed by Chen et \textit{al.} in \cite{laneman} assuming a half-duplex relay. In this paper, the authors focused on the special case of two users, one relay channel and highlighted the significant gains provided by the MAF protocol and the complexity reduction it offers compared to previously proposed protocols. The two-user MAF relay channel is equivalent to a \textit{virtual} multi-antenna MAC (in fact MIMO-MAC, two transmit antennas per user in this case).

In \cite{helmut,Helmut-new}, G\"artner and B\"olcskei presented a detailed analysis of the MAC based on the different error types that can be encountered in this channel and derived the space-time code design criteria for multiantenna MACs. Moreover, they showed that their code design criteria are optimal with respect to the DMT of the channel, \cite{DMTMAC}, and proved that, for a MIMO-MAC, outage analysis allows a rigorous characterization of the dominant error event regions. In other words, outage and error probabilities have the same behaviors. Based on this fundamental result, we presented in a previous work \cite{Nous-ISIT-1} an optimal space-time coding scheme for the MIMO-MAC.

Our main contribution, here, is a new construction of a family of optimal space-time block codes for the two-user, single relay MAF relay channel inspired by the code designed for the MIMO-MAC and based on the DMT of this channel derived in \cite{laneman}. Recall that the users can't cooperate together, hence, without any confusion, in the sequel, the terminology ``cooperative'' and ``non-cooperative'' simply represent channels with and without relay, respectively.

\section{System model}
We use boldface capital letters $\boldsymbol{M}$ to denote matrices. $\mathcal{CN}$ represents the complex Gaussian random variable. [.]$^\top$ (\textit{resp.} [.]$^\dagger$) denotes the matrix transposition (\textit{resp.} conjugated transposition) operation. A $2x \times y$ matrix $\boldsymbol{M}_2$, resulting from the concatenation of the first $x$ rows of the initial $2x \times y$ matrix, $\boldsymbol{M}_1$ and the remaining $x$ rows is denoted (Matlab\textregistered ~notation) \begin{equation*}
 \boldsymbol{M}_2\triangleq [\boldsymbol{M}_1(1:x,1:y) \boldsymbol{M}_1(x+1:2x,1:y)]
\end{equation*}
% The dot equal operator $\doteq$ denotes asymptotic equality in the high SNR regime, \textit{i.e.}, \begin{eqnarray*}
% p_1 \doteq p_2 ~\text{means}~ \lim_{\mathsf{SNR} \rightarrow \infty}\frac{\log p_1}{\log \mathsf{SNR}}=\lim_{\mathsf{SNR} \rightarrow \infty}\frac{\log p_2}{\log \mathsf{SNR}}.
% \end{eqnarray*}

\subsection{The Multiple Access Relay Channel}
\begin{figure}
\centering
\includegraphics[width=0.7\linewidth,keepaspectratio]{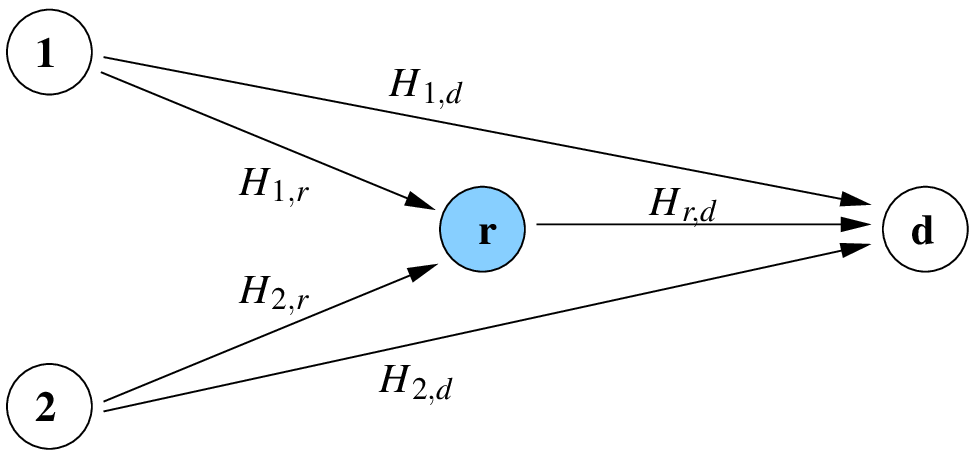}
\caption{A 2-user multiple access relay channel.}
\label{fig:MARC}
\end{figure}
The MAR channel is a MAC with one or more relays helping the users to communicate with the destination while the cooperation between the users is not allowed. In this work, we consider the case of two users single-relay MAR channel. Let $n_t$, $n_r$ and $n_d$ be the number of antennas at the transmitters, the relay and the destination, respectively. The channel model is illustrated in figure \ref{fig:MARC} where $\boldsymbol{H}_{i,d}$, $\boldsymbol{H}_{i,r}$ and $\boldsymbol{H}_{r,d}$ are $n_d \times n_t$, $n_r \times n_t$ and $n_d \times n_r$ independent matrices denoting user $i$-destination, user $i$-relay and relay-destination, respectively, with zero-mean unit variance \textit{i.i.d.} Gaussian entries, \textit{i.e.}, $h_{i,j}$ $\sim$ $\mathcal{CN} (0,1)$. %all equipped with one antenna and one receiver with $n_r$ antennas  For simplicity, we only consider in this work the case where $n_t=n_r=1$ for all $n_r$.

Let $T$ denotes the length of a cooperation frame. All the fading coefficients remain constant within this cooperation frame but change independently from one frame to the other. Moreover, we assume that the Channel State Information (CSI) can be tracked at the receiver, though it is not available at the transmitters. Note that it is assumed that the receiver has knowledge of all CSIs, including those of the user-relay links.

\subsection{Diversity-Multiplexing tradeoff}
The diversity-multiplexing tradeoff (DMT) was introduced and fully characterized in \cite{Zheng-1} in the context of a point to point communication. In this notion, a coding scheme {$\mathcal{C}(\mathsf{SNR})$} is said to achieve \textit{multiplexing gain r} and \textit{diversity gain d} if \begin{eqnarray*}
\lim_{\mathsf{SNR} \rightarrow \infty}\frac{R(\mathsf{SNR})}{\log \mathsf{SNR}}=r~\text{and}~\lim_{\mathsf{SNR} \rightarrow \infty}\frac{P_e(\mathsf{SNR})}{\log \mathsf{SNR}}= - d
\end{eqnarray*} where $R(\mathsf{SNR})$ is the data rate measured by bits per channel use (BPCU) and $P_e(\mathsf{SNR})$ is the average error probability using the maximum likelihood (ML) decoder. The optimal achievable tradeoff $d(r)$ can be found as the exponent of the outage probability in the high $\mathsf{SNR}$ regime.
% The diversity multiplexing tradeoff (DMT) of any linear fading Gaussian channel, $d(r)$, can be found as the exponent of the outage probability in the high $\mathsf{SNR}$ regime, \textit{i.e.}, \begin{equation}
% P_{out}(r \log \mathsf{SNR}) \doteq \mathsf{SNR}^{-d(r)}.
% \end{equation}
Consider the following examples that will have a major importance in the development of this paper.

The DMT of an $n_r \times n_t$ Rayleigh MIMO point-to-point channel is a piecewise-linear function joining points \cite{Zheng-1}\begin{equation}
d_{n_t,n_r}(k)=(n_r-k)(n_t-k),~k=0,1,\dots,\min(n_t,n_r)
\end{equation}

The DMT of the symmetric\footnote{Diversity orders and multiplexing gains per user are identical ($r$ and $d$)} MAC was introduced and fully characterized in \cite{DMTMAC} \begin{equation}\label{eq:dmt_diffr}
\centering
d_{MAC}(r) =
\left\{ \begin{array}{ll}
d_{n_t,n_r}(r), ~~~~~~ r \leq \min(n_t,\frac{n_r}{K+1})\\
d_{Kn_t,n_r}(Kr), ~~~ r \geq \min(n_t,\frac{n_r}{K+1})
\end{array} \right.
\end{equation}

\section{The Multi-Access Amplify-Forward protocol}
In the Multi-Access Amplify-Forward (MAF) presented in \cite{laneman}, both users transmit their informations throughout an entire cooperation frame (two slots). Due to the half-duplex constraint, the relay listens to both users during the first slot, then, in the second slot, it simply amplifies and forward the signal it received. The simplicity of this protocol is its main advantage, in contrast to other protocols, such as the DDF and CF defined previously, that add a significant complexity to the relaying terminal. In addition to the simplicity of the MAF, authors showed that the proposed protocol outperforms both the DDF in the high multiplexing regime and the CF protocol in the low multiplexing regime.

\subsection{Signal model}
The MAF cooperation strategy can be simply illustrated as in the following figure where plain line slots represent transmission mode, whereas dashed slots represent the listening mode.
\begin{figure}[ht]
\centering
\includegraphics[width=0.85\linewidth,keepaspectratio]{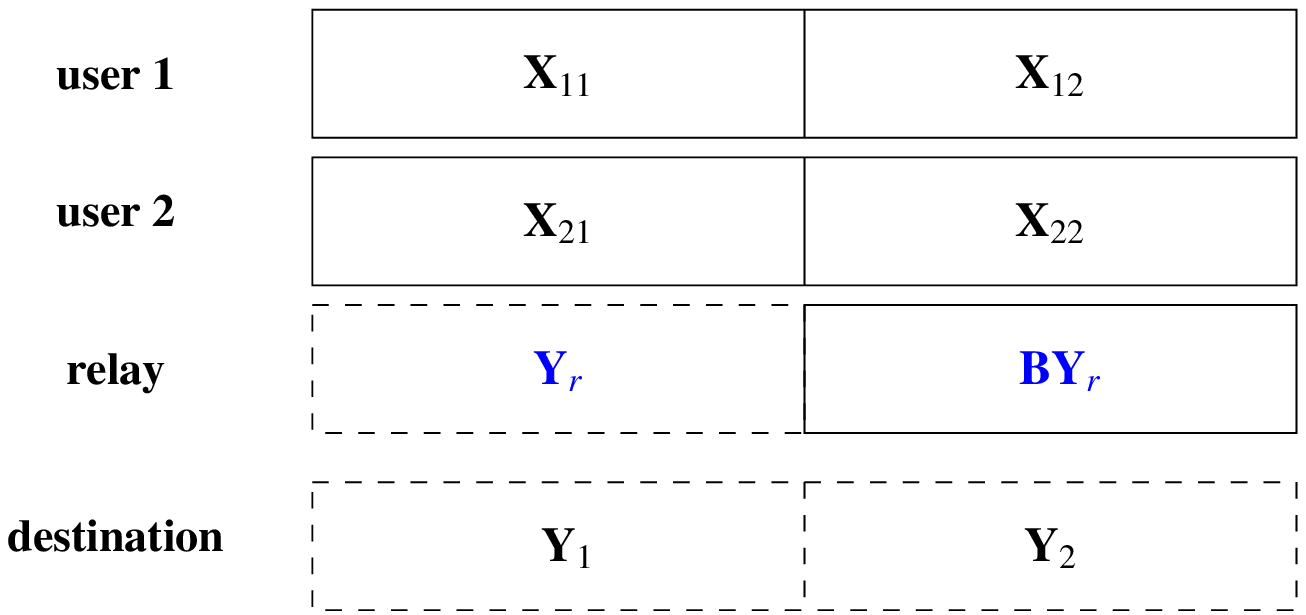}
\caption{A 2-user multiple access relay channel.}
\label{fig:MAF}
\end{figure}

\begin{figure*}[!t]
\normalsize
\begin{equation}
d_{MAR}(r) \leq \min\left\lbrace d_{3 \times n_r}(r), d_{2 \times (n_r+1)}(r), d_{2 \times n_r}(\frac{r}{2}), d_{1 \times (n_r+1)}(\frac{r}{2}) \right\rbrace\label{eq:dmt-gen}
\end{equation}
\hrulefill
\vspace*{4pt}
\end{figure*}

$\boldsymbol{X}_{ij}$ are $n_t \times \frac{T}{2}$ matrices with \textit{i.i.d.} unit variance entries, representing the space-time signal from user $i$ at the $j^{\mathrm{th}}$ slot. $\boldsymbol{Y}_{j}$'s represent the received signal at the destination
\begin{equation}
\left\{ \begin{array}{ll}
\boldsymbol{Y}_{1}=\sum^{2}_{i=1} \sqrt{P_{i1}} \boldsymbol{H}_{i,d} \boldsymbol{X}_{i1} + \boldsymbol{V}_1 \\
\boldsymbol{Y}_{r}=\sum^{2}_{i=1} \sqrt{P_{i1}} \boldsymbol{H}_{i,r} \boldsymbol{X}_{i1} + \boldsymbol{W} \\
\boldsymbol{Y}_{2}=\sqrt{P_{r}}\boldsymbol{H}_{r,d} \boldsymbol{B} \boldsymbol{Y}_{r} + \sum_{i=1}^{2} \sqrt{P_{i2}} \boldsymbol{H}_{i,d} \boldsymbol{X}_{i2} + \boldsymbol{V}_2
\end{array} \right.
\end{equation}
where $\boldsymbol{V}_1$, $\boldsymbol{V}_2$ and $\boldsymbol{W}$ are independent AWGN matrices with normalized \textit{i.i.d.} entries. $\sqrt{P_{ij}}$ and $\sqrt{P_{r}}$ denote user$i$'s transmission power at the $j^{\mathrm{th}}$ slot and the relay's transmission power, respectively. Each power is a fraction ($\pi_{ij}$ and $\pi_{r}$) of the average received $\mathsf{SNR}$ at the destination.\footnote{The total transmit power in both time slots is $(n_t\sum\pi_{ij}+n_r\pi_r)\mathsf{SNR}$. Since the channel coefficients and the AWGN are normalized, $(n_t\sum\pi_{ij}+n_r\pi_r)\mathsf{SNR}$ represents the average received $\mathsf{SNR}$ for both time slots. We choose the values of $\pi$'s satisfying $n_t\sum\pi_{ij}+n_r\pi_r=2$.} $\boldsymbol{B}$ is an $n_d \times n_d$ normalization matrix subject to the power constraint $\mathtt{E}\left\lbrace ||\boldsymbol{B}\boldsymbol{Y}_r||^{2}_{F} \right\rbrace \leq \frac{T}{2}n_r$ \cite{sheng}.
Taking the same footsteps as those in \cite{sheng}, we obtain the following equivalent channel model
\begin{equation}
\boldsymbol{\tilde{Y}}_j=\sum_{i=1}^{2}\boldsymbol{\tilde{H}}_i \boldsymbol{\tilde{X}}_{ij} + \boldsymbol{z}_j,~~~ j=1,\dots,T/2
\end{equation}
where $\boldsymbol{\tilde{X}}_{ij}\triangleq\left[ \boldsymbol{X}_{i1}[j]^{\mathsf{T}}, \boldsymbol{X}_{i2}[j]^{\mathsf{T}}\right]^{\mathsf{T}}$ and $\boldsymbol{\tilde{Y}}_{j}\triangleq\left[ \boldsymbol{Y}_1[j]^{\mathsf{T}}, \boldsymbol{Y}_2[j]^{\mathsf{T}}\right]^{\mathsf{T}}$ are the vectorized transmitted and received signals with $\boldsymbol{M}[i]$ denoting the $i^{\mathrm{th}}$ column of the matrix $\boldsymbol{M}$. $z_i \sim \mathcal{CN} (0,\boldsymbol{I})$ is the equivalent AWGN, the equivalent channel matrix of user $i$ is
\begin{equation}
\boldsymbol{\tilde{H}}_i \triangleq \left[ \begin{matrix}
\sqrt{P_{i1}} \boldsymbol{H}_{i,d} & 0 \\
\sqrt{P_{r}}\sqrt{P_{i1}}\boldsymbol{\Omega}\boldsymbol{H}_{r,d}\boldsymbol{B}\boldsymbol{H}_{i,r} & \sqrt{P_{i2}}\boldsymbol{H}_{i,d}\end{matrix}\right]
\end{equation}
$\boldsymbol{\Omega}$ being the whitening matrix satisfying
$$
(\boldsymbol{\Omega}^{\dagger}\boldsymbol{\Omega})^{-1}=(\boldsymbol{\Omega}\boldsymbol{\Omega}^{\dagger})^{-1}=\boldsymbol{I}+P_r(\boldsymbol{H}_{r,d}\boldsymbol{B})(\boldsymbol{H}_{r,d}\boldsymbol{B})^{\dagger}.
$$

\section{Construction of distributed space-time codes for the MAF relay channel}
In this section, we present our new construction of space-time codes for the MAF relay channel following the same footsteps as the construction of optimal space-time codes for the MIMO Amplify-Forward cooperative channel in \cite{sheng} combined with the construction of codes for the MIMO-MAC \cite{Nous-ISIT-1}. 

\subsection{Structure of the codewords}
We assume that the modulation used by both users is either a quadrature amplitude modulation (QAM) or an hexagonal (HEX) modulation. We denote $\mathbb{P}$ the field $\mathbb{Q}(i)$ (\textit{resp.} $\mathbb{Q}(j)$), representing the modulated symbols. We denote $\mathbb{F}$ a Galois extension of degree $K$ over $\mathbb{P}$ with Galois group $\mathrm{Gal}(\mathbb{F}/\mathbb{P})=\{\tau_1, \tau_2, \dots, \tau_K\}$.  $\mathbb{K}$ is a cyclic extension of degree $n_t$ on $\mathbb{F}$ and $\sigma$ the generator of its Galois group. 
% Let $\gamma\in\mathbb{F}$ such that $\gamma$ is not a norm in $\mathbb{K}$.

Let $\mathcal{X}$ be an optimal space-time code for the MIMO-MAC \cite{Nous-ISIT-1}. We define the mother codeword of user $k$, $\boldsymbol{M}_k$, as a single-user component of a codeword of $\mathcal{X}$ given by \[
\boldsymbol{M}_k=\left[\begin{array}{cccc}
\boldsymbol{\Gamma}\tau_1(\boldsymbol{\Xi}_k) & \boldsymbol{\Gamma}\tau_2(\boldsymbol{\Xi}_k) & \dots & \tau_K(\boldsymbol{\Xi}_k)
\end{array}\right]\]
where $\boldsymbol{\Xi}_k$ is an $2n_t \times 2n_t$ matrix and $\boldsymbol{\Gamma}$ is a multiplication matrix factor for the $k-1$ first matrices of $\boldsymbol{M}_k$. Now consider an equivalent code $\mathcal{C}$ whose codewords (per user) are in the form \[
\boldsymbol{C}_k\triangleq [\boldsymbol{X}_k(1:n_t,1:2Kn_t) \boldsymbol{X}_k(n_t+1:2n_t,1:2Kn_t)]
\]
Then $\mathcal{C}$ achieves the optimal DMT of the $K$-user MAF channel with $n_t$ transmit antennas per user, by transmitting $\boldsymbol{C}_k$ by user $k$, $k=1,\dots,K$ in each cooperation frame. The code $\mathcal{C}$ is of length $T=4Kn_t$.

The key idea of the proof of this result is that, by scaling the size of the underlying QAM constellations by a factor of $\mathsf{SNR}^{r}$, the exponent of $\mathsf{SNR}$ in the asymptotic expression of the error probability varies as the optimal DMT. 

For convenience of presentation, we consider as an example the two-user MAR channel where one single-antenna relay node is assigned to assist the two multiple-access users.
% This work focuses on the special two-user scenario for simplicity of presentation and since the optimal DMT of this channel, used as a performance metric, was derived for this particular case. Each user is assumed to have one transmit antenna.
% Recall that the users are not allowed to help each other and the relay node is constrained by the half-duplex assumption.

\subsection{Two single-antenna-user MAF Relay channel}
\subsubsection*{Optimal DMT}
\begin{figure}[ht]
\centering
\includegraphics[width=0.7\linewidth,keepaspectratio]{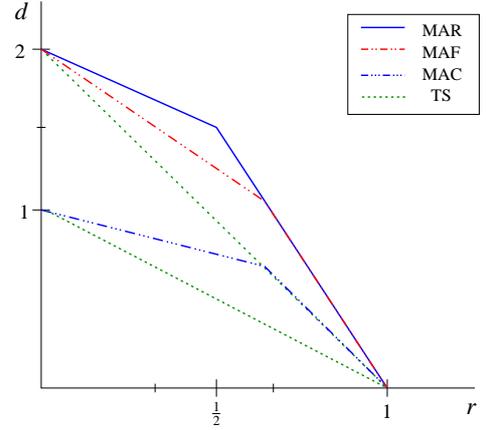}
\caption{DMT of the multi-access relay channel (MAR), multi-access amplify-forward (MAF) relay channel and multiple-access channel (MAC).}
\label{fig:DMT}
\end{figure}

An upper bound on the optimal diversity gain for the symmetric two single-antenna users MAR channel with $n_r$ received antennas is given in (\ref{eq:dmt-gen}) using a simple min-cut max-flow examination of the scheme in figure \ref{fig:MARC} as in \cite{On-the-opt}. Consider as an example the two-user MAR with a single receive antenna. The optimal DMT of this channel is upper bounded by
\begin{equation}
\centering
d_{MAR}(r) \leq
\left\{ \begin{array}{ll}
2-r ,\ \text{for} ,\ 0 \leq r \leq \frac{1}{2} \\
3\left(1-r\right) ,\ \text{for} ,\ \frac{1}{2} \leq r \leq 1
\end{array} \right.
\end{equation}

The DMT of the MAF relay channel was derived in \cite{laneman} \begin{equation}
d_{MAF}(r)=\left\{ \begin{array}{ll}
2-\frac{3}{2}r ,\ \text{for} ,\ 0 \leq r \leq \frac{2}{3} \\
3\left(1-r\right) ,\ \text{for} ,\ \frac{2}{3} \leq r \leq 1
\end{array} \right.
\end{equation}

Figure \ref{fig:DMT} illustrates these two DMTs, the DMT achieved by the MAC obtained using (\ref{eq:dmt_diffr}) for $K=2, n_T=n_r=1$, as well as the DMT achieved by a time-sharing scheme. This comparison reveals the significant advantage that multiple users can potentially gain from a single MAF relay and shows that the MAF protocol achieves the optimal DMT for $2/3 \leq r \leq 1$, \cite{laneman}. It also shows the suboptimality of the time-sharing strategy. Note that, if time-sharing is considered, the system without cooperation is equivalent to a point-to-point scheme. However, with cooperation, the MAF protocol is equivalent to the Non-orthogonal Amplify-Forward (NAF) protocol \cite{on-the-achi}. Our goal is to construct a space-time code that achieves the optimal DMT of the two-user MAF relay channel.

\subsubsection*{Code construction}
\begin{figure*}[!t]
\normalsize
\begin{equation}
\boldsymbol{\Xi}_k=\frac{1}{\sqrt{5}}\left[\begin{array}{cc}
\alpha . (s_{k,1}+s_{k,2}\zeta_8+s_{k,3}\theta+s_{k,4}\zeta_8\theta) & \alpha . (s_{k,5}+s_{k,6}\zeta_8+s_{k,7}\theta+s_{k,8}\zeta_8\theta)\\
\zeta_8 \bar{\alpha} . (s_{k,5}+s_{k,6}\zeta_8+s_{k,7}\bar{\theta}+s_{k,8}\zeta_8\bar{\theta}) & \bar{\alpha} . (s_{k,1}+s_{k,2}\zeta_8+s_{k,3}\bar{\theta}+s_{k,4}\zeta_8\bar{\theta})\end{array}\right]\label{eq:xi}
\end{equation}
\hrulefill
\vspace*{4pt}
\end{figure*}

For the two-user MAF relay channel, we propose the following code. We first design each user's mother codeword. This design aims to construct the Golden code on the base field $\mathbb{Q}(\zeta_8)$ \cite{sheng} instead of the base field $\mathbb{Q}(i)$. Then, using results in \cite{Nous-ISIT-1}, the equivalent mother codeword is designed.

Let $\mathbb{F}=\mathbb{Q}(\zeta_8)$ be an extension of $\mathbb{Q}(i)$ of degree $K=2$, with $\zeta_8=e^{\frac{i \pi}{4}}$ and $\mathbb{K}=\mathbb{F}(\sqrt{5})=\mathbb{Q}(\zeta_8, \sqrt{5})$. Let $\sigma: \theta=\frac{1+\sqrt{5}}{2} \mapsto \bar{\theta}=\frac{1-\sqrt{5}}{2}$, $\alpha=1+i-i\theta$ and $\bar{\alpha}=1+i-i\bar{\theta}$. User's $k$ mother codeword $\boldsymbol{X}_k$ is
\begin{equation}
\boldsymbol{M}_k=\left[\begin{array}{cc}
\boldsymbol{\Xi}_k & \tau(\boldsymbol{\Xi}_k)\end{array}\right]\label{eq:codeword-k}
\end{equation}
with $\boldsymbol{\Xi}_k$ defined by (\ref{eq:xi}) and $\tau$ changes $\zeta_8$ into $-\zeta_8$. $s_{kj}$ denotes the $j^{\mathrm{th}}$ information symbol of user $k$. This code uses 8 QAM symbols per user. Finally, we get the equivalent mother codeword matrix

\begin{equation}\label{eq:codeword-eq}
\boldsymbol{M}=\left[\begin{array}{c}\boldsymbol{M}_1\\\boldsymbol{M}_2\end{array}\right]=\left[\begin{array}{cc}
\boldsymbol{\Xi}_1 & \tau(\boldsymbol{\Xi}_1)\\
\boldsymbol{\Gamma}\boldsymbol{\Xi}_2 & \tau(\boldsymbol{\Xi}_2)\end{array}\right]
\end{equation}
with
\begin{equation*}
\boldsymbol{\Gamma} = \left[\begin{array}{cc}
0 & 1 \\
i & 0
\end{array}\right].
\end{equation*}

The equivalent code $\mathcal{C}$ has codewords (per user) in the form \begin{equation}\label{eq:codeword-k-2}
\boldsymbol{C}_k\triangleq [\boldsymbol{M}_k(1:1,1:4) \boldsymbol{M}_k(2:2,1:4)]
\end{equation}

The proposed code, of length $T=4Kn_t=8$, achieves the optimal DMT of the $(K=2, n_t=1, n_d)$ MAF relay channel by transmitting (for each user $k$), in each cooperation frame, codewords as in (\ref{eq:codeword-k-2}).

\begin{proof} (sketch) If one of the users (say user $2$) is not in error, then the receiver can cancel the signal it receives from this user and the system is equivalent to a single-user single-relay cooperative system. The MAF protocol is, in this case, equivalent to the Non-orthogonal Amplify-and-Forward (NAF), \cite{on-the-achi}. User $1$'s codeword $\boldsymbol{C}_1$ is given in (\ref{eq:codeword-k-2}) and is simply equivalent to the distributed Golden Code of \cite{sheng} which is known to be an DMT achieving space-time block code for the single-relay single-antenna NAF channel.
% \[
% \left[\begin{array}{cc}\boldsymbol{H_1} & 0 \\ 0 & \boldsymbol{H_1} \end{array}\right]
% \left[\begin{array}{cc}\boldsymbol{\Xi_1} & 0 \\0 & \tau(\boldsymbol{\Xi_1})\end{array}\right]
% \]

If both users are in error, the mother codewords are those of \cite{Nous-ISIT-1}.
% The matrix $\boldsymbol{\Gamma}$ should be carefully chosen in order to the code to be DMT achieving.
% We choose the unitary matrix, element of $\mathbb{Q}(i,\sqrt{5})$, given in (\ref{eq:gamma}). 
Thus, the proposed code is DMT achieving.

% Note that there is two ways of extending $\mathbb{Q}(i)$ to $\mathbb{K}=\mathbb{Q}(\zeta_8, \sqrt{5})$ as follows \begin{figure}[ht]
% \noindent \begin{centering}
% \includegraphics[width=0.4\columnwidth,keepaspectratio]{extension}
% \par\end{centering}
% \caption{Extending $\mathbb{Q}(i)$ up to $\mathbb{K}$}
% \end{figure}
\end{proof}

\section{Numerical results}
In this section, we present the numerical results obtained by Monte-Carlo simulations. We assume that the power is allocated equally so that no a-priori advantage is given to any link over another one. We first compare the outage performance of the MAF relay channel to the time-sharing strategy where the channel is shared among the users in an orthogonal multiple-access manner. We also consider the non-cooperative scenario in both multiple-access and time-sharing cases to highlight the benefit of the relay. The performance of the proposed scheme is then measured by the word error rate (WER) \textit{vs} received $\mathsf{SNR}$ and compared to the time sharing scheme.

At the receiver side, we use (when needed) an MMSE-DFE preprocessing combined with lattice decoding as a way to tackle the problem of the rank deficiency resulting from $n_d$ being smaller than $K \times n_t$. In \cite{ElGamal-search} it is shown that an appropriate combination of left, right preprocessing and lattice decoding, yields significant saving in complexity with very small degredation with respect to the ML performance.
% More precisely, left preprocessing (forward filter, $\boldsymbol{F}$) modifies the channel matrix and the noise vector such that the resulting closest lattice point search has a much better conditioned channel matrix. Moreover, right preprocessing (backward filter, $\boldsymbol{B}$) is used to change the lattice basis such that it becomes more convenient for the search stage.

\begin{figure}[ht]
\noindent \begin{centering}
\includegraphics[scale=0.49,angle=270]{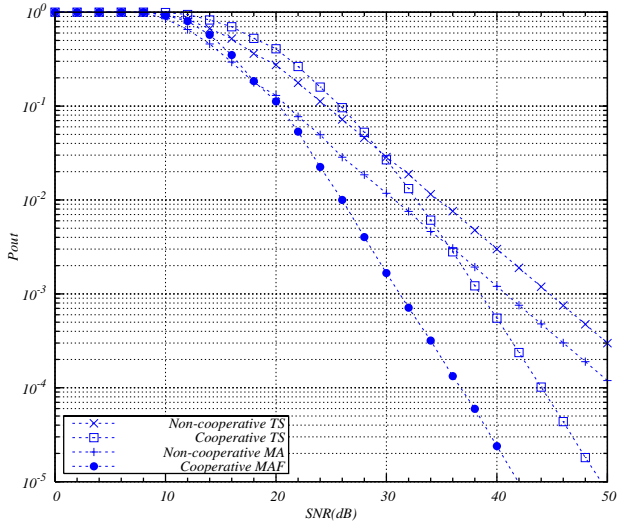}
\par\end{centering}
\caption{\label{fig:po-2}Outage performance of two-user MAR channel, $n_r=1$, MAF protocol \emph{vs} time sharing, $R=2 BPCU$.}
\end{figure}

\begin{figure}[ht]
\noindent \begin{centering}
\includegraphics[scale=0.49,angle=270]{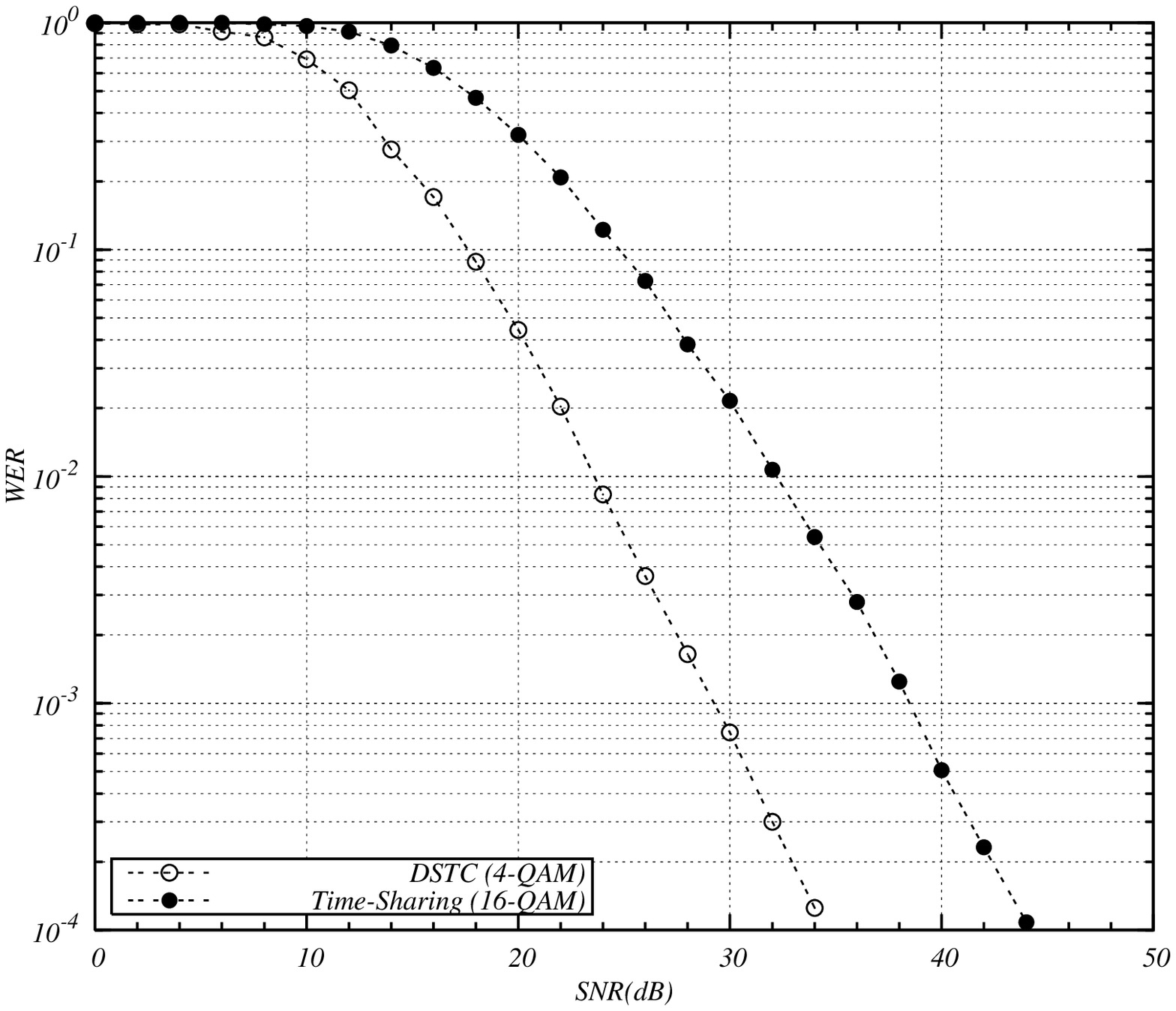}
\par\end{centering}
\caption{\label{fig:po-4}Performance of the Space-Time Code designed for the two-user MAF relay channel, $n_r=1$, 4-QAM.}
\end{figure}

% \subsection{Outage and code performance}
Outage performance of the two-user channel is illustrated in figures \ref{fig:po-2} (with $n_d=1$) and \ref{fig:po-4} (with $n_d=2$) for different spectral efficiencies, $2-$ and $4-$ BPCU, respectively. Both non-cooperative and cooperative systems are considered for two different access strategies: simultaneous transmission (multi-access MA) and orthogonal access (time-sharing TS). This study gives the theoretical limit of the channel in each scenario and will be used as an analysis tool to evaluate the performance of coding schemes.

Two important observations can be made: In the high $\mathsf{SNR}$ regime, as in the single user case, the existence of a relay helping the users to reach the destination offers a significant gain in both multi-access and time-sharing schemes. Compared to the time-sharing scenario, the multi-access scenario achieves the same diversity order but offers a significant gain, in both cooperative and non-cooperative channels, that increases with the spectral efficiency.

\begin{figure}[ht]
\noindent \begin{centering}
\includegraphics[scale=0.49,angle=270]{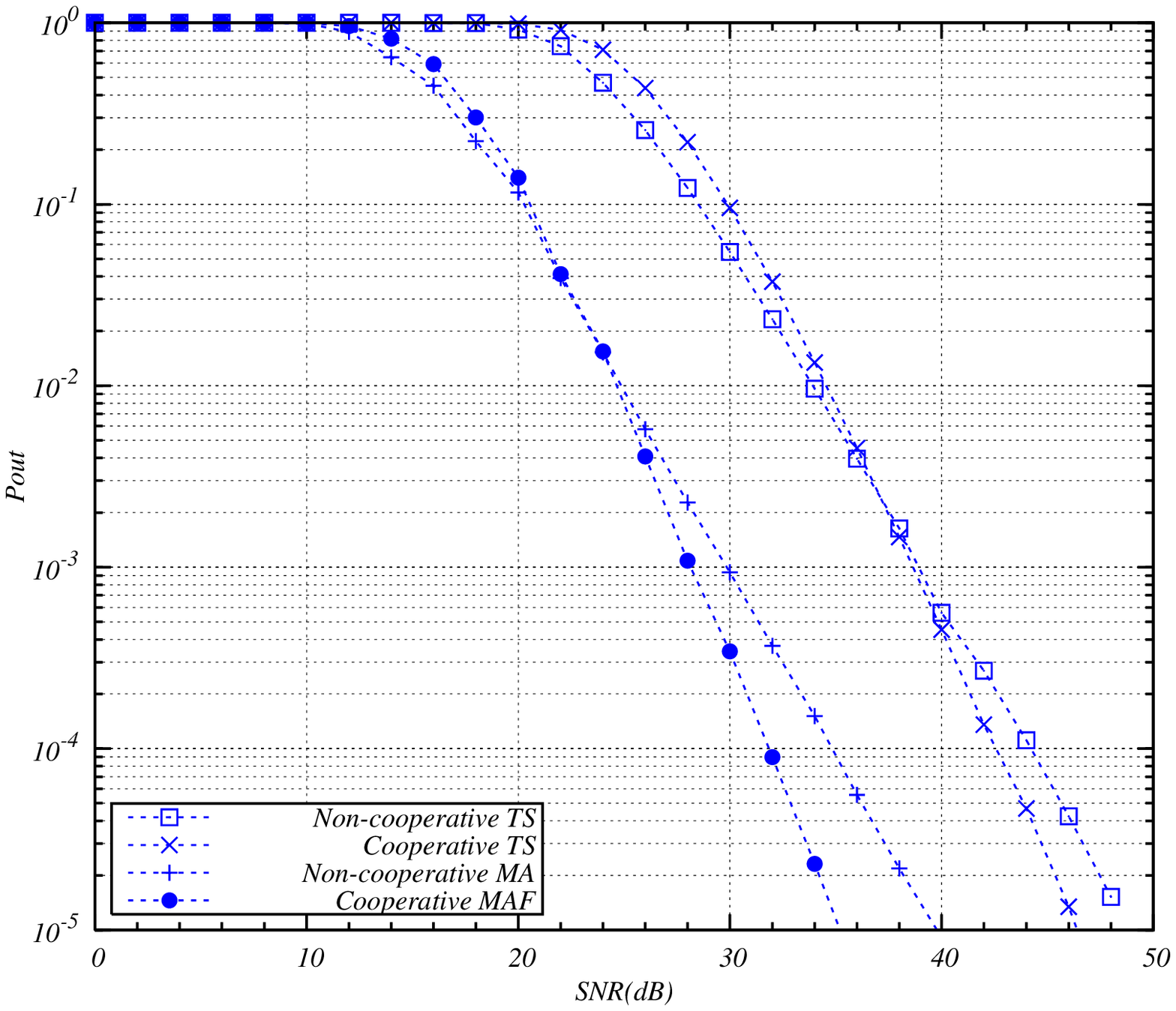}
\par\end{centering}
\caption{\label{fig:pe-4}Outage performance of two-user MAR channel, $n_r=2$, MAF protocol \emph{vs} time sharing, $R=4 BPCU$.}
\end{figure}

\begin{figure}[ht]
\noindent \begin{centering}
\includegraphics[scale=0.49,angle=270]{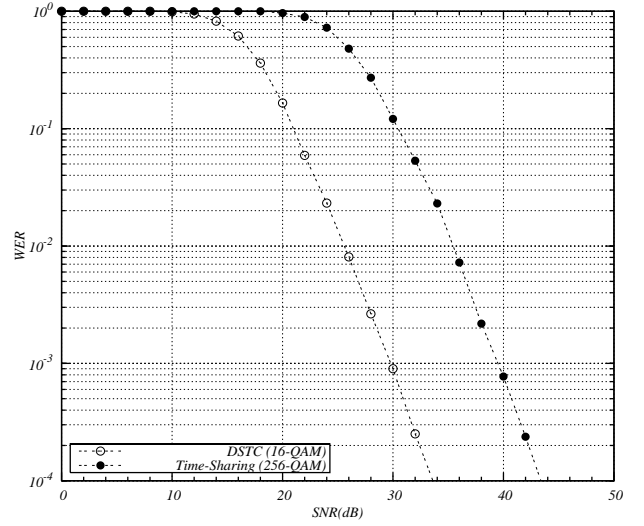}
\par\end{centering}
\caption{\label{fig:pe-16}Performance of the Space-Time Code designed for the two-user MAF relay channel, $n_r=2$, 16-QAM.}
\end{figure}

% \subsection{Code performance}
Now, we consider the new coding scheme whose performance is shown in figures \ref{fig:pe-4} and \ref{fig:pe-16}. In the first one, the number of receive antennas is set to $1$, whereas in the second one it is set to $2$ in order to simplify the decoding. Compared to the time-sharing scheme, the proposed code achieves the same diversity order (2 for $n_d=1$ and 3 for $n_d=2$) but offers a significant performance gain that depends on the spectral efficiency of the system. In order to highlight this dependence on $R$, users information symbols are carved from different QAM constellations, \textit{e.g.} 4-QAM and a 16-QAM for the coded scheme (16-QAM and 256-QAM, repectively, for the time-sharing scheme). At WER = $10^{-3}$, a gain of $8$ dB is observed when a 4-QAM constellation is considered. When we increase the spectral efficiency (16-QAM), this gain increases to $10$ dB. Interestingly, if we compare the performance of the coded scheme to the outage performance of the channel, the same behavior can be observed.


\section{Conclusion}
In this paper, the multiple access relay channel with no channel side information at the transmitters is studied. We considered the multi-access amplify-forward relaying protocol which experiences low complexity at the relay and achieves the optimal diversity-multiplexing tradeoff for $\frac{2}{3} \leq r \leq 1$. In order to get advantage of the gain offered by the existence of the relay compared to a system without relaying and by multi-access technique compared to the time-sharing, we propose a new construction of distributed space-time block codes. In addition to its practical encoding and decoding interest, the new coding scheme achieves the optimal DMT of the $2$-user MAF relay channel. Simulation results show that the new codes offer a significant performance gain compared to the time sharing scheme.

\end{document}